\documentclass[conference]{IEEEtran}
\IEEEoverridecommandlockouts
\usepackage{cite}
\usepackage{amsmath,amssymb,amsfonts}
\usepackage{algorithmic}
\usepackage{graphicx}
\usepackage{textcomp}
\usepackage{xcolor}

\usepackage{filecontents}

\def\BibTeX{{\rm B\kern-.05em{\sc i\kern-.025em b}\kern-.08em
    T\kern-.1667em\lower.7ex\hbox{E}\kern-.125emX}}
\begin{document}

\title{Feature Imitating Networks Enhance the Performance, Reliability and Speed of
Deep Learning on Biomedical Image Processing Tasks\\

}

\author{\IEEEauthorblockN{Shangyang Min}
\small
\IEEEauthorblockA{\textit{Computer Science} \\
\textit{Brown University}\\
Providence, RI, United States \\
shangyang\_min@brown.edu}
\and
\IEEEauthorblockN{Hassan B. Ebadian}
\IEEEauthorblockA{\textit{Radiation Oncology} \\
\textit{Henry Ford Hospital}\\
Detroit, MI, USA \\
hbagher1@hfhs.org}
\and
\IEEEauthorblockN{Tuka Alhanai}
\IEEEauthorblockA{\textit{Computer Engineering} \\
\textit{New York University}\\
Abu Dhabi, UAE \\
tuka.alhanai@nyu.edu}
\and
\IEEEauthorblockN{Mohammad M. Ghassemi}
\IEEEauthorblockA{\textit{Computer Science} \\
\textit{Michigan State University}\\
East Lansing, MI, USA \\
ghassem3@msu.edu}
}

\maketitle
\begin{abstract}
Feature-Imitating-Networks (FINs) are neural networks that are first trained to approximate closed-form statistical features (e.g. Entropy), and then embedded into other networks to enhance their performance. In this work, we perform the first evaluation of FINs for biomedical image processing tasks. We begin by training a set of FINs to imitate six common radiomics features, and then compare the performance of larger networks (with and without embedding the FINs) for three experimental tasks: COVID-19 detection from CT scans, brain tumor classification from MRI scans, and brain-tumor segmentation from MRI scans. We found that models embedded with FINs provided enhanced performance for all three tasks when compared to baseline networks without FINs, even when those baseline networks had more parameters. Additionally, we found that models embedded with FINs converged faster and more consistently compared to baseline networks with similar or greater representational capacity. The results of our experiments provide evidence that FINs may offer state-of-the-art performance for a variety of other biomedical image processing tasks.
\end{abstract}

\section{Introduction}

\begin{figure}[t!]
\centering
\includegraphics[width=\columnwidth, trim={3cm 1cm 1.5cm 0cm}, clip]{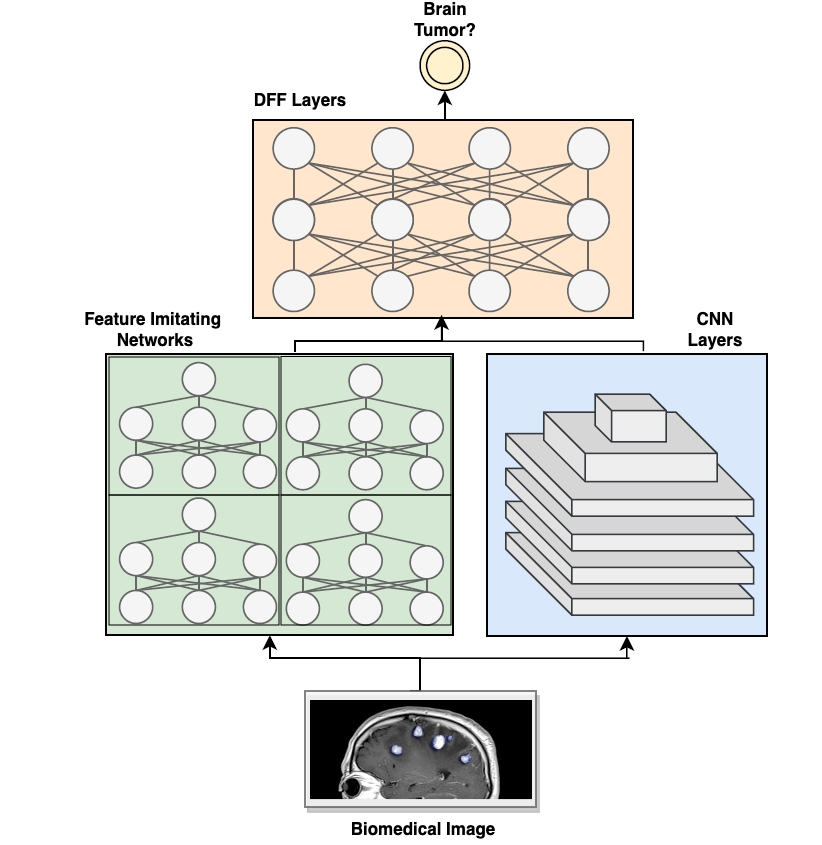} 
\caption{An illustration of our approach that integrates Feature Imitating Networks (FINs) into larger model structures for biomedical image processing tasks. A biomedical image (bottom) is passed to a CNN (blue), and a set of FINs (green), pre-trained to approximate a set of closed-form radiomics features (see Section \ref{sec:Meth}). The results from FINs and the CNN layers are received by a fully connected DFNN (orange) to predict the outcome. The representations learned by the FINs evolve during fine-tuning for the task.}
\label{fig:modelfin}
\end{figure} 

\subsection{Motivation}
\noindent Medical imaging has greatly benefitted from the rise of Deep Learning techniques for tasks including image segmentation, classification, detection, retrieval, reconstruction, filtering, denoising, and super-resolution \cite{8657645}. Indeed, there are countless studies that have demonstrated the superiority of Deep Learning approaches compared to classical machine learning algorithms (e.g. Support Vector Machines) using expert characteristics \cite{8241753}. Importantly, however, there remain domains where Deep Learning approaches have yet to provide breakthrough performance: when data is too small \cite{whang2023data}, computational resources needed to train models are too large \cite{desislavov2023trends}, or where model interpretation is a strict requirement \cite{menghani2023efficient, vellido2020importance}. This is often the case in medical imaging tasks; a recent review article on biomedical image classification techniques by Tchapga \textit{et al.} reported that Neural Networks were less accurate, less tolerant to redundant attributes, slower to train, and more likely to overfit compared to SVM models with expert features on a variety of biomedical image classification tasks when data was scarce \cite{tchito2021biomedical}. 

\noindent Feature Imitating Networks (FIN) refers to a recently developed Deep Learning paradigm that solves the aforementioned challenges of standard Deep Learning in data-scarce environments \cite{saba2022feature}. A FIN is a neural network pre-trained to approximate one or more closed-form statistical features that are thought to be relevant for a given task. For example, assume that the heterogeneity of image texture is useful for a task; A FIN might be pre-trained to approximate Shannon's entropy. After being trained, the FINs would be integrated within a larger, more complex network architecture that inherits the feature representation of Entropy (See Figure \ref{fig:modelfin}). This integration bypasses the need to learn the feature from (large) data and also overcomes the representational rigidity that would result from including the closed-form feature as an input to the model, directly. That is, as part of network fine-tuning, the representation captured by the FIN evolves from the static feature representation it was first trained to emulate into a specific instantiation of that feature that is most effective for the task at hand; for example, a FIN that is designed to emulate Shannon’s entropy may evolve into a Tsalis entropy representation during the fine-tuning process.

\subsection{Our contributions}
Previously, FIN-embedded models have shown state-of-the-art performance on several biomedical signal processing tasks using EEG \cite{saba2022feature}. However, there has been no prior work to evaluate the potential of FINs for biomedical image processing tasks. In this work, we extend the FIN paradigm to enable their emulation of common radiomics features and apply them to three biomedical imaging tasks: COVID-19 detection from CT scans, brain tumor classification from Magnetic Resonance Imaging (MRI), and brain-tumor segmentation from MRI. We demonstrate that FINs provide best-in-class performance for all three tasks, while converging faster and more consistently compared to networks with similar or greater capacity.

 \subsection{Related work}

In the following subsections, we present an overview of relevant literature on biomedical image classification and segmentation, both of which are relevant to the experimental tasks in this study. More specifically, we provide a brief discussion of the approaches that will serve as baselines for the experimental results discussed later in this paper.

\subsubsection{Image classification}
The state-of-the-art for most biomedical image classification tasks use Deep Learning (CNNs, specifically) when large datasets are available \cite{puttagunta2021medical}, or SVMs with expert features when datasets are smaller \cite{wagner2021radiomics, zhang2023radiomics}. The use of Deep Learning methods (including CNNs) will continue to grow as dataset sizes grow and become better integrated \cite{SABA201914, ibrahim2021radiomics}. Hence, we compared our FIN-integrated models, against several CNN architectures as well as an SVM for image classification tasks.

\subsubsection{Segmentation}
UNets are reported to provide state-of-the-art performance for biomedical image segmentation tasks \cite{kaur2021ga}. UNets are neural networks designed for image segmentation that
contain two parts: (1) a CNN architecture and (2) an up-sampling path. Using highly limited training samples, UNets can create highly detailed segmentation maps of images. UNets are also reported to require less training time than alternative segmentation models \cite{9446143}. Hence, we compare our FIN-integrated models against the UNet architecture for the segmentation task.

\section{METHODS} \label{sec:Meth}
In this section, we provide a discussion of our proposed approach. More specifically, we discuss the procedure used to select and train the FINs which were later applied to the three experimental tasks outlined in section \ref{sec:Exp}. For our purposes, the FINs were first trained to imitate several common radiomics features used in medical imaging tasks; these FINs were then integrated within a larger network architecture and fine-tuned for the specific task (See Figure \ref{fig:modelfin}). The code used to train the FINs can be found online\footnote{https://github.com/HAAIL/FINs-for-biomedical-imaging}.

\subsection{Radiomics Features Selected} There are three categories of features that are often used for radiomics tasks: first-order features (e.g. entropy, skewness), shape-based features (e.g mesh surface), and texture-based features (e.g. autocorrelation) \cite{KUMAR20121234}. For the purposes of this study, we trained six FINs to emulate common radiomics features. Five FINs were each independently trained to emulate texture-based features: Autocorrelation\cite{moradmand2020impact}, Gray Level Variance \cite{simpson2020impact}, Cluster Shade \cite{gabelloni2020can}, Difference Entropy \cite{ogbonnaya2021prediction}, and Size Zone Non-uniformity \cite{wang2020radiomics}. The sixth FIN was trained to imitate a first-order feature: Skewness\cite{coppola2021heterogeneity}. We intentionally excluded shape-based features because the extraction of shape features requires significantly more computational power (over 10x) than texture or first-order features.

\subsection{Data and Training Approach} 

Using PyRadiomics \cite{van2017computational}, we first computed radiomics features on images from TCIA\cite{lungCT}. Next, for each feature, a separate FIN was trained to imitate the feature, given the image.  Each FIN was a deep feed-forward neural network (DFNN). For each FIN, we used a $relu$ as the activation function for feature approximation. The training was conducted on a \textit{NVIDIA A100 GPU}. Several topological configurations were explored at random and the best configuration was retained (see the online repository to access the final FIN for each feature). Following training, the FINs were integrated within the baseline network for the experimental tasks in Section III.

\section{Experiments} \label{sec:Exp}
In this section, we describe three radiomics tasks (two classification, and one image segmentation) where we compare the performance of FIN-integrated models against baseline approaches. Within the description of each experiment, we elaborate on the data used for the task, as well as any fine-tuning considerations for integration of the FINs within the baseline approaches for the task. All experiments described here used publicly available data. The code used to regenerate the experimental results can be found online.

\subsection{Experiment I}\label{sec:Exp1}
The task for the first experiment was binary classification of COVID-19 using Lung CT images. 

\paragraph*{Data and Prepossessing} The data for the experiment consisted of $8,439$ Lung CT scans from the COVID-19 Lung CT dataset \cite{covid19binary}. The lung scans consisted of $944$ non-COVID images and $7,496$ COVID images. The images were portioned into 10-folds; in each fold, 90\% of the data was used for model training and validation, and the remaining 10\% of the data was used for model testing. All models were assessed according to the mean and standard deviation of their Area Under the Receiver Operator Characteristic Curve (AUROC) across the ten folds, as well as the number of epochs required for convergence of the validation loss.

\begin{table}[!b]
\centering
\small
\caption{Comparison of the AUROC (mean and standard deviation) as well as the average epochs until convergence for our proposed approach (FINs), and the three baseline approaches (SVM, DFNN, CNN) across the 10-folds in experiment I.}
\resizebox{\columnwidth}{!}{%
\begin{tabular}{|c|c|c|c|}
\hline
     & AUROC ($\mu$) & AUROC ($\sigma$)   & \begin{tabular}[c]{@{}l@{}}Training\\ Epochs ($\mu$)\end{tabular}  \\ \hline
SVM & 0.611    & 0.0259   & N/A    \\ \hline
DFNN & 0.667    & 0.0629   & 6.9    \\ \hline
CNN  & 0.995    & 0.0050   & 7.2    \\ \hline
FINs  & 0.998    & 0.0029   & 5.7    \\ \hline
\end{tabular}%
}

\label{table:exp1}
\end{table}

\begin{table*}[t!]
\centering
\small
\caption{Mean and standard deviation of F-1 scores and accuracy, as well as the number of epochs until converge of each model by repeating the experiment \ref{sec:Exp2}, and compare the results of our proposed approach (row 3) against the two baseline approaches.}
\resizebox{\textwidth}{!}{%
\begin{tabular}{|l|c|c|c|c|c|}
\hline
                & F-1 Score  ($\mu$) & F-1 score ($\sigma$) & Accuracy ($\mu$) & Accuracy ($\sigma$) & Training epochs ($\mu$) \\ \hline
RGB CNN         & 0.617       & 0.0177  & 0.677      & 0.0177     & 4.25                     \\ \hline

Grey-scaled CNN & 0.629      & 0.0137     & 0.684      & 0.0089     & 4.38                    \\ \hline

FINs            & 0.643        & 0.0084      & 0.697    & 0.0059     & 4.25                     \\ \hline
\end{tabular}%
}
\label{table:exp2}
\end{table*}

\paragraph*{Models} We trained three binary classification models using the collected data: (1) an SVM with a polynomial kernel and a soft-margin tolerance parameter set to 1, embodying the traditional approach to feature engineering in machine learning, (2) a DFNN using the radiomics features as inputs, (3) a CNN model that utilized the raw images as inputs with a final DFNN layer, and (4) a CNN model with an embedded FIN ensemble that imitated the six radiomics features (described in section \ref{sec:Meth}).

The FINs were trained to imitate the radiomics features of an entire image using a separate open source imaging data set from the lung CT segmentation challenge \cite{lungCT} provided by TCIA \cite{clark2013cancer}. The challenge dataset consisted of $9,593$ CT images from $60$ thoracic patients undergoing radiation treatment. To ensure a fair comparison between the models, we explored 20 topological configurations; the best performing topological configuration was retained.  

Importantly, we explored CNN and DFNN baseline configurations with the same (or greater) number of parameters as the FIN-embedded model; this ensured that the raw representational capacity of the models was comparable. The number of parameters for the final CNN, DFNN, and FIN ensemble models was 149M, 150M, and 147M respectively.

\paragraph*{Results}
In Table \ref{table:exp1} we compare the performance of the three models with respect to the mean AUROC across the ten testing folds, the standard deviation of the test set performance, and the number of epochs required for the models to converge. As seen in the table, the SVM had the lowest overall AUROC (0.611) of the four models, the DFNN using the raw radiomics features had the lowest AUROC (0.667) of the three Deep Learning network approaches. The mean AUROC of the FIN-embedded model was only slightly better than the CNN model (0.998 and 0.995 respectively); however, the standard deviation of the FIN-embedded model was 42\% lower than the CNN approach and 95\% lower than the DFNN. Importantly, the FIN-embedded model required the fewest number of epochs before convergence; the FIN-embedded model converged 20\% faster than the CNN, and faster 17\% than the DFNN. These results imply that FIN-embedded models provide enhanced classification performance, which is more robust and faster to train than the alternative approaches.

\subsection{Experiment II}\label{sec:Exp2}
The task for the second experiment was a multiclass brain tumor classification. Our second experiment builds on the encouraging results seen in Experiment I (Section \ref{sec:Exp1}) using the same FIN and the structure of the CNN model baseline.

\paragraph*{Data and Prepossessing} The data for this experiment consisted of brain MRI scans from a brain tumor classification dataset \cite{brainMulti}. Training data consisted of $2,870$ scans, describing four outcome classes: glioma tumors ($n=826$), meningioma tumors ($n=822$), pituitary tumors ($n=827$), and non-tumors ($n=395$). Within the test set, there were $100$ scans for glioma tumors, $115$ scans for meningioma tumors, $74$ scans for pituitary tumors, and $105$ scans for non-tumors. Before modeling, all images were converted to dimensions of $512 \times 512$ using \textit{SimpleITK} \cite{yaniv2018simpleitk}, and grayscaled using the following formula: \textit{$0.2989 \times R + 0.5870 \times G + 0.1140 \times B$}.

\paragraph*{Models} We trained three multiclass classification models with the same structure as the baseline CNN and FINs in experiment I (Section \ref{sec:Exp1}) using the collected data: (1) a CNN model with an embedded FIN ensemble that imitated the six radiomics features described in section \ref{sec:Meth}, (2) a baseline CNN model with RGB image inputs, and (3) a baseline CNN model with grey-scaled image inputs. 

We trained two CNN models, one with grey-scaled inputs and the other one with RGB inputs to ensure that there was no color sensitivity when modeling with CNNs. The number of parameters for the two CNNs and FIN ensemble was 149M and 147M respectively. We shuffled and regrouped 90\% of the data for model training, and the remaining 10\% of the data was used for validation. All models were assessed according to the mean and standard deviation of their F-1 score, accuracy, and the number of epochs until convergence of the validation loss.

\paragraph*{Result} In Table \ref{table:exp2}, we compare the performance of our proposed approach with the baseline models. We observed that the CNN model was more sensitive to grey-scaled inputs as demonstrated by a higher average accuracy and F-1 score, and lower standard deviation, relative to the RGB model. Importantly, the FINs ensemble outperformed both the RGB and grey-scaled CNN models: the FINs ensemble had a higher F-1 score and accuracy, with a 39\% lower standard deviation for the F-1 score and 33\% lower standard deviation for accuracy than the best performing baseline model (grey-scaled CNN). The three models all converged at similar epochs in terms of the validation loss.

\subsection{Experiment III} \label{sec:Exp3}
The task for the third experiment was a segmentation task using brain MRI scans from TCIA and TCGA.

\paragraph*{Data and Prepossessing} The data for this experiment consisted of $3,929$ brain MRI scans with corresponding segmentation masks \cite{brainMRI}. The images were divided into three categories: 70\% for training, 15\% for validation, and 15\% for testing. All images were normalized from the range of 0-255 to 0-1.

\paragraph*{Models} We trained two models using the training data: (1) A standard UNet with $64$ filters that increased quadratically to $1,024$ filters, followed by up-sampling \cite{9446143}, (2) a standard UNet with FINs imitating the six features described in section \ref{sec:Meth}. The FINs in this task were trained using the data from this task to imitate the radiomics features in $256$ sub-segments of the image. The outputs of the FINs were inserted after the UNet's first max-pooling layer. All models were evaluated according to their intersection over union (IoU) and dice similarity coefficient: standard measures of image segmentation accuracy \cite{jha2020kvasir}.

\begin{figure}[t]
\centering
\includegraphics[width=1.02\columnwidth]{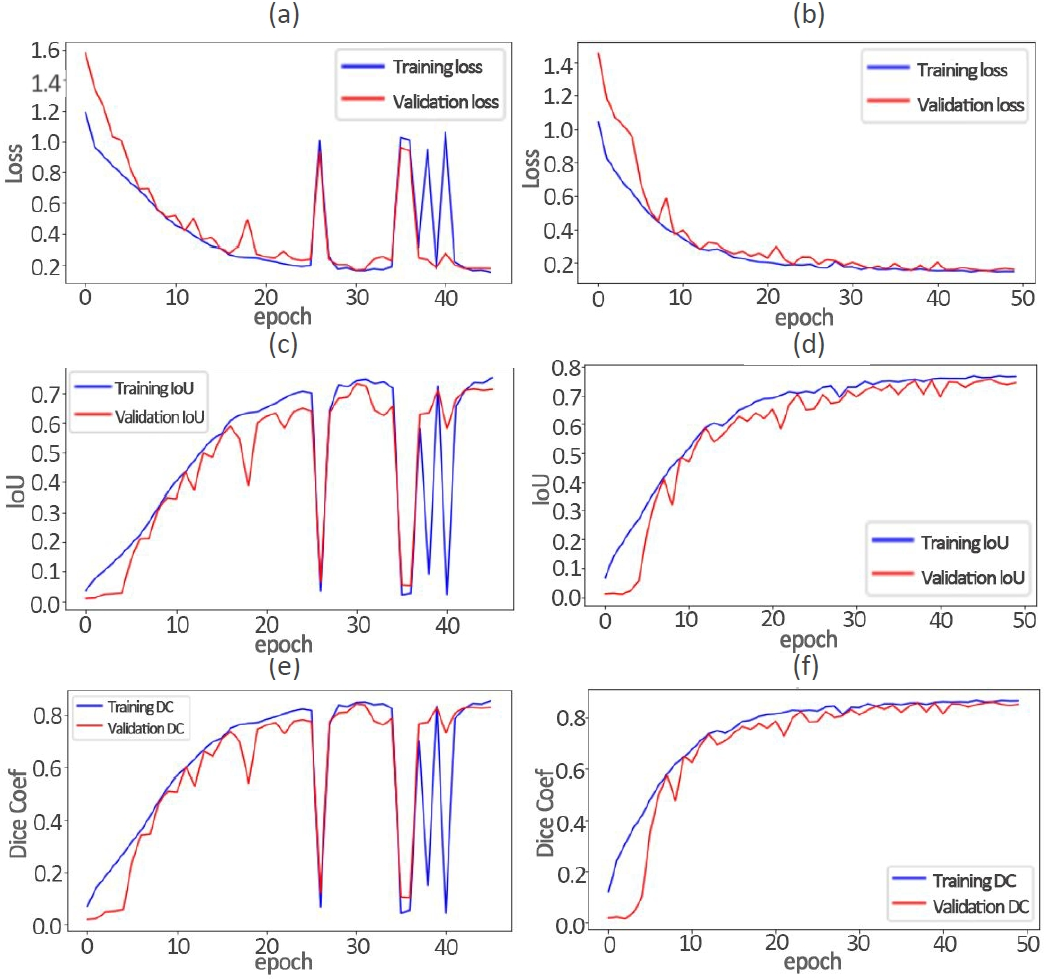} 
\caption{UNet training performance before (left column) and after (right column) insertion of FINs. Plots show loss (row 1), IoU (row 2), and dice coefficient (row 3) during the $50$ epochs of training: (a) loss of basic UNet,(b) loss of UNet with FINs, (c) IoU of basic UNet, (d) IoU of UNet with FINs, (e) dice coefficient of basic UNet and (f) dice coefficient of UNet with FINs. The red line represents the validation set loss metrics while the blue line represents for training set loss metrics. Overall training with FINs is observed to be more stable.}
\label{fig:exp3}
\end{figure}

\paragraph*{Result} Figure \ref{fig:exp3} compares the loss, IoU, and dice similarity coefficient of the two models (UNet, UNet + FINs) during the training and validation phases. As seen in the figure, the baseline model of the UNet approach experienced problems more frequently with loss instability and poor validation set performance. In contrast, the FIN approach reduced the sudden irregularities in loss, and the corresponding effects on the IoU and dice coefficient measures. When evaluated in the test set, the performance of UNet combined with FINs was higher than that of the UNet model alone, as demonstrated by an increase in both IoU (from $0.72$ to $0.74$), and dice coefficient (from $0.835$ to $0.851$).

\section{Discussion} \label{sec:Disc}


\paragraph*{Key Results} The results of all three experiments provide evidence that embedding FINs within conventional neural networks enhances their ability to learn faster and more reliably without additional parameters. More specifically, the results of experiment I (see Table \ref{table:exp1}), demonstrate that FINs can provide enhanced performance on \textit{binary classification tasks}; the results of experiment II (see Table \ref{table:exp2}) demonstrate that FINs provide enhanced performance on \textit{multinomial classification tasks}; the results of experiment III (see Figure \ref{fig:exp3}) demonstrate that FINs provide enhanced performance on \textit{segmentation tasks}. Critically, in all three experiments, FIN-embedded models required fewer training epochs for validation loss convergence and exhibited lower variance over multiple folds compared to conventional architectures (CNN or DFNN) with similar representational power (i.e. number of parameters).

Critically, the enhanced performance of  FIN-embedded models is not achieved by simply making the raw feature values available to the networks. In Experiment I, our results indicated that simply using the features in a traditional machine learning model underperformed the FIN-integrated method. We also found that the DFNN model, with access to the raw features that the FINs were trained to imitate, performed more poorly than a basic CNN model for radiomics tasks. This highlights that FINs are able to \textit{adapt} the features to be best suited for the task at hand; that is, using a FIN is not equivalent to simply passing in feature values at the input of the network. 

An important attribute of the FIN models is not only their higher performance, but also the speed of model training (fewer epochs), and stability of the model performance (lower variance in test performance across folds); these properties of FINs may be explained by the fact that they are imitating features which are known to be important for the task at hand. That is, the network is initiated in a parameter space which we have reason to believe is better than a random initialization.

\paragraph*{Future Directions}
An obvious direction for future research is to generate a larger variety of radiomics FINs beyond the six used in this study. Furthermore, it would be interesting to experiment with the cross-functionality of FINs for tasks in different domains.

\section*{Acknowledgments}
This research was generously supported by the National Science Foundation (Award 2320050), and an HFH+MSU Health Sciences Center Cancer Grant (Award IP00545511). T.A. acknowledges the support of the NYUAD Center for AI and Robotics. We thank the data generated by the TCGA Research Network.

\bibliographystyle{IEEEtran}
\bibliography{refs}

\begin{thebibliography}{10}
\providecommand{\url}[1]{#1}
\csname url@samestyle\endcsname
\providecommand{\newblock}{\relax}
\providecommand{\bibinfo}[2]{#2}
\providecommand{\BIBentrySTDinterwordspacing}{\spaceskip=0pt\relax}
\providecommand{\BIBentryALTinterwordstretchfactor}{4}
\providecommand{\BIBentryALTinterwordspacing}{\spaceskip=\fontdimen2\font plus
\BIBentryALTinterwordstretchfactor\fontdimen3\font minus \fontdimen4\font\relax}
\providecommand{\BIBforeignlanguage}[2]{{%
\expandafter\ifx\csname l@#1\endcsname\relax
\typeout{** WARNING: IEEEtran.bst: No hyphenation pattern has been}%
\typeout{** loaded for the language `#1'. Using the pattern for}%
\typeout{** the default language instead.}%
\else
\language=\csname l@#1\endcsname
\fi
#2}}
\providecommand{\BIBdecl}{\relax}
\BIBdecl

\bibitem{8657645}
M.~Hatt, C.~Parmar, J.~Qi, and I.~El~Naqa, ``Machine (deep) learning methods for image processing and radiomics,'' \emph{IEEE Transactions on Radiation and Plasma Medical Sciences}, vol.~3, no.~2, pp. 104--108, 2019.

\bibitem{8241753}
J.~Ker, L.~Wang, J.~Rao, and T.~Lim, ``Deep learning applications in medical image analysis,'' \emph{IEEE Access}, vol.~6, pp. 9375--9389, 2018.

\bibitem{whang2023data}
S.~E. Whang, Y.~Roh, H.~Song, and J.-G. Lee, ``Data collection and quality challenges in deep learning: A data-centric ai perspective,'' \emph{The VLDB Journal}, vol.~32, no.~4, pp. 791--813, 2023.

\bibitem{desislavov2023trends}
R.~Desislavov, F.~Mart{\'\i}nez-Plumed, and J.~Hern{\'a}ndez-Orallo, ``Trends in ai inference energy consumption: Beyond the performance-vs-parameter laws of deep learning,'' \emph{Sustainable Computing: Informatics and Systems}, vol.~38, p. 100857, 2023.

\bibitem{menghani2023efficient}
G.~Menghani, ``Efficient deep learning: A survey on making deep learning models smaller, faster, and better,'' \emph{ACM Computing Surveys}, vol.~55, no.~12, pp. 1--37, 2023.

\bibitem{vellido2020importance}
A.~Vellido, ``The importance of interpretability and visualization in machine learning for applications in medicine and health care,'' \emph{Neural computing and applications}, vol.~32, no.~24, pp. 18\,069--18\,083, 2020.

\bibitem{tchito2021biomedical}
C.~Tchito~Tchapga, T.~A. Mih, A.~Tchagna~Kouanou, T.~Fozin~Fonzin, P.~Kuetche~Fogang, B.~A. Mezatio, and D.~Tchiotsop, ``Biomedical image classification in a big data architecture using machine learning algorithms,'' \emph{Journal of Healthcare Engineering}, vol. 2021, 2021.

\bibitem{saba2022feature}
S.~Saba-Sadiya, T.~Alhanai, and M.~M. Ghassemi, ``Feature imitating networks,'' in \emph{ICASSP 2022-2022 IEEE International Conference on Acoustics, Speech and Signal Processing (ICASSP)}.\hskip 1em plus 0.5em minus 0.4em\relax IEEE, 2022, pp. 4128--4132.

\bibitem{puttagunta2021medical}
M.~Puttagunta and S.~Ravi, ``Medical image analysis based on deep learning approach,'' \emph{Multimedia Tools and Applications}, vol.~80, no.~16, pp. 24\,365--24\,398, 2021.

\bibitem{wagner2021radiomics}
M.~W. Wagner, K.~Namdar, A.~Biswas, S.~Monah, F.~Khalvati, and B.~B. Ertl-Wagner, ``Radiomics, machine learning, and artificial intelligence—what the neuroradiologist needs to know,'' \emph{Neuroradiology}, pp. 1--11, 2021.

\bibitem{zhang2023radiomics}
T.~Zhang, T.~Tan, R.~Samperna, Z.~Li, Y.~Gao, X.~Wang, L.~Han, Q.~Yu, R.~G. Beets-Tan, and R.~M. Mann, ``Radiomics and artificial intelligence in breast imaging: a survey,'' \emph{Artificial Intelligence Review}, vol.~56, no. Suppl 1, pp. 857--892, 2023.

\bibitem{SABA201914}
\BIBentryALTinterwordspacing
L.~Saba, M.~Biswas, V.~Kuppili, E.~{Cuadrado Godia}, H.~S. Suri, D.~R. Edla, T.~Omerzu, J.~R. Laird, N.~N. Khanna, S.~Mavrogeni, A.~Protogerou, P.~P. Sfikakis, V.~Viswanathan, G.~D. Kitas, A.~Nicolaides, A.~Gupta, and J.~S. Suri, ``The present and future of deep learning in radiology,'' \emph{European Journal of Radiology}, vol. 114, pp. 14--24, 2019. [Online]. Available: \url{https://www.sciencedirect.com/science/article/pii/S0720048X19300919}
\BIBentrySTDinterwordspacing

\bibitem{ibrahim2021radiomics}
A.~Ibrahim, S.~Primakov, M.~Beuque, H.~Woodruff, I.~Halilaj, G.~Wu, T.~Refaee, R.~Granzier, Y.~Widaatalla, R.~Hustinx \emph{et~al.}, ``Radiomics for precision medicine: Current challenges, future prospects, and the proposal of a new framework,'' \emph{Methods}, vol. 188, pp. 20--29, 2021.

\bibitem{kaur2021ga}
A.~Kaur, L.~Kaur, and A.~Singh, ``Ga-unet: Unet-based framework for segmentation of 2d and 3d medical images applicable on heterogeneous datasets,'' \emph{Neural Computing and Applications}, vol.~33, no.~21, pp. 14\,991--15\,025, 2021.

\bibitem{9446143}
N.~Siddique, S.~Paheding, C.~P. Elkin, and V.~Devabhaktuni, ``U-net and its variants for medical image segmentation: A review of theory and applications,'' \emph{IEEE Access}, vol.~9, pp. 82\,031--82\,057, 2021.

\bibitem{KUMAR20121234}
\BIBentryALTinterwordspacing
V.~Kumar, Y.~Gu, S.~Basu, A.~Berglund, S.~A. Eschrich, M.~B. Schabath, K.~Forster, H.~J. Aerts, A.~Dekker, D.~Fenstermacher, D.~B. Goldgof, L.~O. Hall, P.~Lambin, Y.~Balagurunathan, R.~A. Gatenby, and R.~J. Gillies, ``Radiomics: the process and the challenges,'' \emph{Magnetic Resonance Imaging}, vol.~30, no.~9, pp. 1234--1248, 2012, quantitative Imaging in Cancer. [Online]. Available: \url{https://www.sciencedirect.com/science/article/pii/S0730725X12002202}
\BIBentrySTDinterwordspacing

\bibitem{moradmand2020impact}
H.~Moradmand, S.~M.~R. Aghamiri, and R.~Ghaderi, ``Impact of image preprocessing methods on reproducibility of radiomic features in multimodal magnetic resonance imaging in glioblastoma,'' \emph{Journal of applied clinical medical physics}, vol.~21, no.~1, pp. 179--190, 2020.

\bibitem{simpson2020impact}
G.~Simpson, J.~C. Ford, R.~Llorente, L.~Portelance, F.~Yang, E.~A. Mellon, and N.~Dogan, ``Impact of quantization algorithm and number of gray level intensities on variability and repeatability of low field strength magnetic resonance image-based radiomics texture features,'' \emph{Physica medica}, vol.~80, pp. 209--220, 2020.

\bibitem{gabelloni2020can}
M.~Gabelloni, L.~Faggioni, S.~Attanasio, V.~Vani, A.~Goddi, S.~Colantonio, D.~Germanese, C.~Caudai, L.~Bruschini, M.~Scarano \emph{et~al.}, ``Can magnetic resonance radiomics analysis discriminate parotid gland tumors? a pilot study,'' \emph{Diagnostics}, vol.~10, no.~11, p. 900, 2020.

\bibitem{ogbonnaya2021prediction}
C.~N. Ogbonnaya, X.~Zhang, B.~S. Alsaedi, N.~Pratt, Y.~Zhang, L.~Johnston, and G.~Nabi, ``Prediction of clinically significant cancer using radiomics features of pre-biopsy of multiparametric mri in men suspected of prostate cancer,'' \emph{Cancers}, vol.~13, no.~24, p. 6199, 2021.

\bibitem{wang2020radiomics}
T.~Wang, J.~Deng, Y.~She, L.~Zhang, B.~Wang, Y.~Ren, J.~Wu, D.~Xie, X.~Sun, and C.~Chen, ``Radiomics signature predicts the recurrence-free survival in stage i non-small cell lung cancer,'' \emph{The Annals of thoracic surgery}, vol. 109, no.~6, pp. 1741--1749, 2020.

\bibitem{coppola2021heterogeneity}
F.~Coppola, M.~Mottola, S.~Lo~Monaco, A.~Cattabriga, M.~A. Cocozza, J.~C. Yuan, C.~De~Benedittis, D.~Cuicchi, A.~Guido, F.~L. Rojas~Llimpe \emph{et~al.}, ``The heterogeneity of skewness in t2w-based radiomics predicts the response to neoadjuvant chemoradiotherapy in locally advanced rectal cancer,'' \emph{Diagnostics}, vol.~11, no.~5, p. 795, 2021.

\bibitem{van2017computational}
J.~J. Van~Griethuysen, A.~Fedorov, C.~Parmar, A.~Hosny, N.~Aucoin, V.~Narayan, R.~G. Beets-Tan, J.-C. Fillion-Robin, S.~Pieper, and H.~J. Aerts, ``Computational radiomics system to decode the radiographic phenotype,'' \emph{Cancer research}, vol.~77, no.~21, pp. e104--e107, 2017.

\bibitem{lungCT}
Y.~J., S.~G., V.~H., V.~E. W., D.~A., L.~T., and G.~M., ``Data from lung ct segmentation challenge (version 3) [data set],'' 2017.

\bibitem{covid19binary}
\BIBentryALTinterwordspacing
M.~Aria, M.~Ghaderzadeh, F.~Asadi, and R.~Jafari, ``Covid-19 lung ct scans,'' 2021. [Online]. Available: \url{https://www.kaggle.com/dsv/1875670}
\BIBentrySTDinterwordspacing

\bibitem{clark2013cancer}
K.~Clark, B.~Vendt, K.~Smith, J.~Freymann, J.~Kirby, P.~Koppel, S.~Moore, S.~Phillips, D.~Maffitt, M.~Pringle \emph{et~al.}, ``The cancer imaging archive (tcia): maintaining and operating a public information repository,'' \emph{Journal of digital imaging}, vol.~26, no.~6, pp. 1045--1057, 2013.

\bibitem{brainMulti}
\BIBentryALTinterwordspacing
S.~Bhuvaji, A.~Kadam, P.~Bhumkar, S.~Dedge, and S.~Kanchan, ``Brain tumor classification (mri),'' 2020. [Online]. Available: \url{https://www.kaggle.com/dsv/1183165}
\BIBentrySTDinterwordspacing

\bibitem{yaniv2018simpleitk}
Z.~Yaniv, B.~C. Lowekamp, H.~J. Johnson, and R.~Beare, ``Simpleitk image-analysis notebooks: a collaborative environment for education and reproducible research,'' \emph{Journal of digital imaging}, vol.~31, no.~3, pp. 290--303, 2018.

\bibitem{brainMRI}
\BIBentryALTinterwordspacing
P.~N., F.~A. E., S.~L., T.~Mikkelsen, E.~J. M., H.~B., S.~V., B.-S. J., and O.~Q., ``The cancer genome atlas low grade glioma collection (tcga-lgg) (version 3) [data set],'' 2016. [Online]. Available: \url{https://www.cancerimagingarchive.net/collection/tcga-lgg/}
\BIBentrySTDinterwordspacing

\bibitem{jha2020kvasir}
D.~Jha, P.~H. Smedsrud, M.~A. Riegler, P.~Halvorsen, T.~d. Lange, D.~Johansen, and H.~D. Johansen, ``Kvasir-seg: A segmented polyp dataset,'' in \emph{International Conference on Multimedia Modeling}.\hskip 1em plus 0.5em minus 0.4em\relax Springer, 2020, pp. 451--462.

\end{thebibliography}

\end{document}